\documentclass{aa}

\usepackage{graphicx}
\usepackage{float}
\usepackage{txfonts}
\usepackage{hyperref}

\newcommand{\parallelsum}{\mathbin{\!/\mkern-5mu/\!}}

\graphicspath{{images/}}

\begin{document} 

   \title{Physics-based model of the adaptive-optics-corrected\\ point spread function \thanks{Our Python codes will be available on \url{https://gitlab.lam.fr/lam-grd-public} as soon as the paper is published.}}
   \titlerunning{Physics based model of the AO-corrected PSF}
   \subtitle{Applications to the SPHERE/ZIMPOL and MUSE instruments}
   \author{R. JL. F{\'e}tick\inst{1,2}\and
          T. Fusco\inst{1,2}\and
          B. Neichel\inst{1}\and
          L. M. Mugnier\inst{2}\and
          O. Beltramo-Martin\inst{1}\and\\
          A. Bonnefois\inst{2}\and
          C. Petit\inst{2}\and
          J. Milli\inst{3}\and
          J. Vernet\inst{4}\and
          S. Oberti\inst{4}\and
          R. Bacon\inst{5}
          }

   \institute{Aix Marseille Univ, CNRS, CNES, LAM, Marseille, France\\
              \email{romain.fetick@lam.fr}
         \and
            ONERA, The French Aerospace Lab BP72, 29 avenue de la Division Leclerc, 92322 Chatillon Cedex, France
        \and
        European Southern Observatory (ESO), Alonso de C{\'o}rdova 3107, Vitacura, Casilla 19001,Santiago, Chile
        \and
        ESO, European Southern Observatory, Karl-Schwarzschild Str. 2, 85748 Garching bei Muenchen, German
        \and
            CRAL, Observatoire de Lyon, CNRS, Universit{\'e} Lyon 1, 9 Avenue Ch. Andr{\'e}, F-69561 Saint Genis Laval Cedex, France
             }

   \date{Received May $3^{rd}$ 2019; accepted May $28^{th}$ 2019}

  \abstract
   {Adaptive optics (AO) systems greatly increase the resolution of large telescopes, but produce complex point spread function (PSF) shapes, varying in time and across the field of view. The PSF must be accurately known since it provides crucial information about optical systems for design, characterization, diagnostics, and image post-processing.}
   {We develop here a model of the AO long-exposure PSF, adapted to various seeing conditions and any AO system. This model is made to match accurately both the core of the PSF and its turbulent halo.}
   {The PSF model we develop is based on a parsimonious parameterization of the phase power spectral density, with only five parameters to describe circularly symmetric PSFs and seven parameters for asymmetrical ones. Moreover, one of the parameters is the Fried parameter $r_0$ of the turbulence's strength. This physical parameter is an asset in the PSF model since it can be correlated with external measurements of the $r_0$, such as phase slopes from the AO real time computer (RTC) or site seeing monitoring.}
   {We fit our model against end-to-end simulated PSFs using the OOMAO tool, and against on-sky PSFs from the SPHERE/ZIMPOL imager and the MUSE integral field spectrometer working in AO narrow-field mode. Our model matches the shape of the AO PSF both in the core and the halo, with a relative error smaller than 1\% for simulated and experimental data. We also show that we retrieve the $r_0$ parameter with sub-centimeter precision on simulated data. For ZIMPOL data, we show a correlation of $97\%$ between our $r_0$ estimation and the RTC estimation. Finally, MUSE allows us to test the spectral dependency of the fitted $r_0$ parameter. It follows the theoretical $\lambda^{6/5}$ evolution with a standard deviation of $0.3$ cm. Evolution of other PSF parameters, such as residual phase variance or aliasing, is also discussed.}
  {}


   \maketitle

\section{Introduction}

Optical systems suffer from aberrations and diffraction effects that limit their imaging performance. For ground-based observations, the point spread function (PSF) is dramatically altered by the atmospheric turbulence that distorts the incoming wavefront \citep{Roddier1981}. The resolution under typical conditions of a seeing-limited telescope does not exceed the diffraction limit of an $\sim 12$ cm aperture. Modern and future large telescopes thus include adaptive optics (AO) systems \citep{Roddier1999} that compensate for the atmospheric turbulence thanks to wavefront sensors and deformable mirrors. The aberrated wavefront is partially corrected and telescopes may operate near their diffraction limited regime. Nevertheless the AO correction is limited by technical issues such as sensor noise, limited number of actuators, or loop delay \citep{Martin2017,Rigaut1998}. This results in a peculiar shape of the PSF made of a sharp peak due to the partial AO correction, and a wide halo caused by the residual turbulence above the AO cutoff frequency.\\

The PSF thus provides critical information about an optical system regarding its preliminary design, calibrations, testings, or diagnostic \citep{Ascenso2015,Ragland2018}. Image post-processing, such as deconvolution \citep{mugnier2004}, also requires knowledge of the PSF. Deconvolution of long-exposure images using parametric PSFs has already been demonstrated in \citet{Drummond1998} and \citet{Fetick2019}. A fine model of the PSF is necessary. The substantial advantage of parametric PSFs is to compress all the important information of the physical PSF into a small number of parameters. The numerical values of these parameters might then be used for comparisons, correlations, or any statistical analysis. Moreover if the PSF parameters are correlated to physical values (e.g. turbulence strength, wind speed, AO residual phase variance), it is possible to better constrain these parameters or better understand the AO response to given observing conditions. We state that an efficient AO PSF model should fulfil the following requirements:
\begin{itemize}
    \item \textbf{Accuracy}. The model must represent accurately the shape of the AO-corrected PSF, especially the two areas corresponding to its central peak and to its wide turbulent halo. The requested accuracy depends on the application of the PSF (e.g. fitting, deconvolution, turbulence monitoring).
    \item \textbf{Versatility and robustness}. The model must be used on different AO system, with different AO correction levels, for different turbulent strengths.
    \item \textbf{Simplicity}. The model must have as few parameters as possible without damaging its versatility or accuracy.
    \item \textbf{Physical parameters}. Such parameters have a physical meaning related to the observing conditions. These parameters have physical units.
\end{itemize}

The literature already provides some models of AO-corrected PSF \citep{Drummond1998,zieleniewski2013} often based on Gaussian, Lorentzian, and/or \citet{moffat1969} models. A trade-off is always drawn between a simple model with few parameters but imprecise, or a more precise but also more complex model. The difficulty often comes from the description of the turbulent halo with only a few parameters. Moreover, to the best of our knowledge, these PSF models rely only on mathematical parameters without direct physical meaning or units.\\

We propose a long-exposure PSF model for AO-corrected telescopes that describes accurately the shape of the PSF; this model is made of a small number of parameters with physical meaning whenever possible. Our method does not parameterize the PSF directly in the focal plane, but rather from the phase power spectral density (PSD). Indeed \citet{goodman1968fourier} and \citet{Roddier1981} have shown that the phase PSD contains all the necessary information to describe the long-exposure atmospheric PSF. Working in the PSD domain allows us to include physical parameters. Then  Fourier transforms give the resulting PSF in the focal plane. Our PSF model also includes pupil diffraction effects or any of the system static aberrations, provided they have been previously characterized.\\

In Sect. \ref{sec:PSFmodel} we first recall the expression of the Moffat function and show its limits for AO PSF description. Then we develop our PSF model, partially based on this Moffat function. Section 3 validates the model by fitting PSFs from numerical simulations and from observations made on two Very Large Telescope (VLT) instruments. Finally Sect. \ref{sec:ccl} concludes our work and discusses direct and future applications for our PSF model.

\section{Description of the PSF model}
\label{sec:PSFmodel}

In the whole paper, we define $(x_R,y_R)$ the reference coordinates that are respectively the detector horizontal and vertical coordinates. We also define $(x,y)$ the proper PSF coordinates (e.g. along the major and minor PSF elongation axis), rotated by an angle $\theta_R$ with respect to the reference frame. The reference frame to PSF frame transformation can be written as
\begin{equation}
    \begin{pmatrix}
    x\\y
    \end{pmatrix}
    =
    \begin{pmatrix}
    \cos\theta_R & \sin\theta_R \\
    -\sin\theta_R & \cos\theta_R
    \end{pmatrix}
    \begin{pmatrix}
    x_R\\y_R
    \end{pmatrix}
.\end{equation}
For the sake of simplicity we will use mainly the rotated coordinates, but it is important to keep in mind that $\theta_R$ is a crucial PSF parameter. We also simplify the notations $x=x(x_R,y_R,\theta_R)$ and $y=y(x_R,y_R,\theta_R)$.\\

In this section we first recall the usual Moffat PSF model, since it encompasses and generalizes Lorentzian and Gaussian models. We demonstrate the advantages of the Moffat model, but also its limitations. This motivates our search for a better PSF model. However, the mathematical expression of the Moffat function will still be used inside our more complete PSF model.

\subsection{Review of the usual Moffat PSF model}
\label{sec:moffat}

The AO-corrected PSF exhibits a sharp corrected peak, with wide wings extension. The \citet{moffat1969} model is often used due to its good approximation of the AO PSF sharp peak \citep{Andersen2006,Muller2006,Davies2012,Orban2015,Rusu2016}. The Moffat function, of amplitude $A$, is written as
\begin{equation}
\label{eq:moffat}
M_A(x,y) = \frac{A}{(1+x^2/\alpha_x^2+y^2/\alpha_y^2)^\beta}
,\end{equation}
with $\alpha_x$, $\alpha_y,$ and $\beta$ strictly positive real numbers. Moreover the condition $\beta>1$ is imposed to ensure a finite integral of the function on the plane. This model encompasses the two-dimensional Lorentzian function for $\beta=1$ and the two-dimensional Gaussian function for $\beta\to +\infty$. The variable $\beta$ parameter thus makes the Moffat function a generalization of Lorentzian and Gaussian ones. Since the PSF has a unit energy, demonstration in Appendix \ref{sec:appendix_moffat} shows that the Moffat multiplicative constant is
\begin{equation}
    A = \frac{\beta-1}{\pi\alpha_x\alpha_y}
,\end{equation}
so the PSF, called $h$, is made of only four free parameters $\alpha_x$, $\alpha_y$, $\beta,$ and $\theta_R$. The Moffat PSF model thus is re-written as
\begin{equation}
    h(x,y) = \frac{\beta-1}{\pi\alpha_x\alpha_y} M_1(x,y) 
,\end{equation}
where the notation $M_1$ must be understood as the Moffat $M_A$ with a multiplicative factor $A=1$.\\

\begin{figure}[!ht]
   \centering
   \includegraphics[width=0.9\columnwidth]{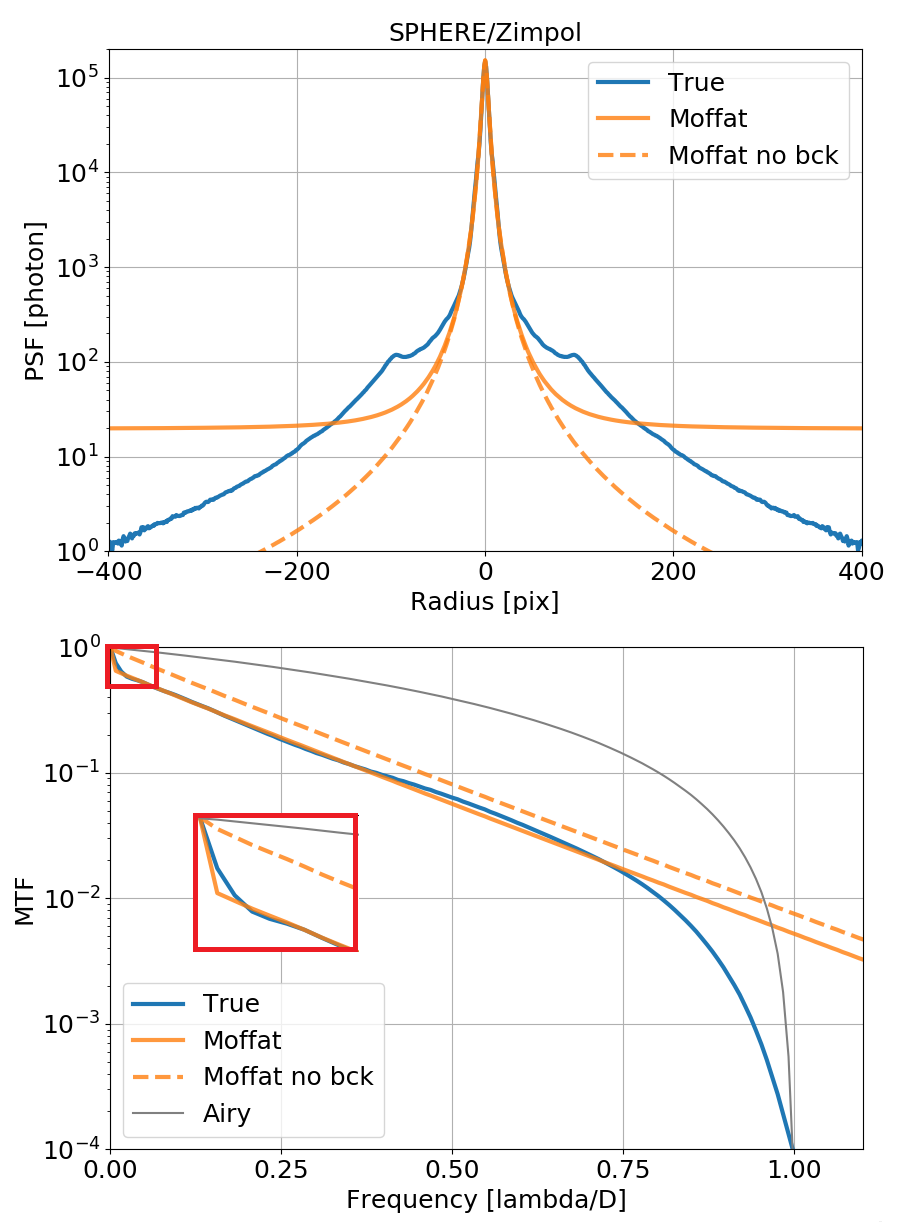}
   \caption{Fitting of a SPHERE/ZIMPOL PSF (blue) using a Moffat model with background (green) and without background (dashed green). Top: PSF, Bottom: MTF. The insert plot is a zoom on the low spatial frequencies.}
\label{fig:ZIMPOL_moffat}
\end{figure}

The full fitting method will be presented in Sect. \ref{sec:validation}, but we show here a preliminary result using the Moffat function to motivate our search for better functions. Indeed, as shown in Fig. \ref{fig:ZIMPOL_moffat}, the Moffat function accurately fits the central peak of an actual PSF, but poorly describes the turbulent halo. Since this halo may contain an important proportion of the PSF energy, depending on the quality of the AO correction, it is necessary to model it accurately. Adding a constant background to the model artificially improves the fitting (lower residuals). However, this method is not suitable since it poorly describes the halo and mistaking the halo for a background will yield the non-physical result of a PSF with an infinite integral on an unlimited field of view. The modulation transfer functions (MTF, bottom plot in Fig. \ref{fig:ZIMPOL_moffat}), which is the modulus of the PSF Fourier transform, also shows that the Moffat does not match well the very low frequencies (halo) and does not model the telescope cutoff frequency. Similarly, none of the static aberrations of the telescope are taken into account. A more physical PSF model than a Moffat is thus required.

\subsection{Image formation theory}

Our PSF model is based on equations of image formation from the phase PSD to the focal plane. Indeed \citet{Roddier1981} has shown that the Fourier transform of the PSF, the optical transfer function (OTF), can be written as the product of the telescope aberrations OTF and the atmospheric turbulent OTF,
\begin{equation}
\label{eq:OTF_tel_atm}
\tilde{\mathrm{h}}(\vec{\rho}/\lambda) = \tilde{\mathrm{h}}_T(\vec{\rho}/\lambda) \cdot \tilde{\mathrm{h}}_A(\vec{\rho}/\lambda)
,\end{equation}
where $\lambda$ is the observation wavelength, $\tilde{h}$ the total OTF, $\tilde{\mathrm{h}}_T$ the telescope OTF, and $\tilde{\mathrm{h}}_A$ the atmospheric OTF. This OTF splitting equation is valid under the hypothesis of a spatially stationary phase. This is the case for a purely turbulent phase, and a good approximation for an AO-corrected phase \citep{conan1994etude}. To establish this result, Roddier also used the fact that the phase distribution follows a Gaussian process, as the sum of a large number of independent turbulent layers. The telescope OTF is simply given by the autocorrelation of the pupil transmission function, whereas the atmospheric OTF is written as
\begin{equation}
\label{eq:OTFatmo}
\tilde{\mathrm{h}}_A(\vec{\rho}/\lambda) = e^{-B_\phi(\vec{0})}e^{B_\phi(\vec{\rho})}
,\end{equation}
with $B_\phi$ the phase autocorrelation function defined as
\begin{equation}
B_\phi(\vec{\rho}) = \langle\langle \phi(\vec{r},t)\phi(\vec{r}+\vec{\rho},t) \rangle_t\rangle_r
.\end{equation}

The Wiener-Khintchine theorem states that the PSD is the Fourier transform of the autocorrelation as
\begin{equation}
    W_\phi(\vec{f}) = \mathcal{F}\{B_\phi(\vec{\rho})\}
,\end{equation}
where $W_\phi$ denotes the phase PSD and $\mathcal{F}$ the Fourier operator; $\vec{f}$ and $\vec{\rho}$ are the Fourier conjugated variables. If we call $\vec{u}$ the angular variable conjugated to $\vec{\rho}/\lambda$, the PSF is written as
\begin{equation}
\label{eq:PSF}
    h(\vec{u}) = \mathcal{F}^{-1} \left\{\tilde{\mathrm{h}}_T(\vec{\rho}/\lambda)~  e^{-B_\phi(\vec{0})}~ e^{\mathcal{F}^{-1}\{W_\phi(\vec{f})\}} \right\}
.\end{equation}
We note that $B_\phi(\vec{0})$ is the residual phase variance and is equal to the integral of $W_\phi$ on the whole frequency plane. This equation shows that only knowledge of the pupil and the static aberrations of the term $\tilde{\mathrm{h}}_T$, and the phase PSD $W_\phi$ are necessary for the description of the long-exposure PSF. Diffraction effects -- such as finite aperture, central obstruction, and spiders -- only depend on the pupil geometry and are known. Static aberrations are second-order effects that can be either neglected (as we show in Sects. \ref{sec:ZIMPOL} and \ref{sec:muse}), or measured \citep{Ndiaye2013} and then included in the $\tilde{\mathrm{h}}_T$ term for more accuracy. The term $\tilde{\mathrm{h}}_T$ being fully determined, now we only have to parameterize the residual phase PSD to model the PSF.\\

\subsection{Parameterization of the phase PSD}
\label{sec:PSDmodel}

Actuators controlling the deformable mirror are separated by a pitch that sets the maximal spatial frequency of the phase that can be corrected by the AO system. This is called the AO spatial cutoff frequency, defined by $f_\text{AO}\simeq N_\text{act}/2D$, where $N_\text{act}$ is the linear number of actuators and $D$ the pupil diameter. This technical limitation induces the peculiar shape of the AO residual phase PSD (and \textit{in fine} a peculiar shape of the PSF). The residual phase PSD is thus separated into two distinct areas:
\begin{itemize}
    \item AO-corrected frequencies $f\leq f_\text{AO,}$
    \item AO-uncorrected frequencies $f>f_\text{AO.}$
\end{itemize}
The uncorrected area is not affected by the AO system, and the phase PSD consequently follows the Kolmogorov law,
\begin{equation}
    W_{\phi,\text{Kolmo}}(\vec{f}) = 0.023 r_0^{-5/3} f^{-11/3} ~~ \text{ , for } f>f_\text{AO}
,\end{equation}
where $r_0$ is the Fried parameter scaling the strength of the turbulence. The halo is thus set by the knowledge of only this $r_0$ parameter.\\

\begin{figure}[!ht]
   \centering
   \includegraphics[width=.9\columnwidth]{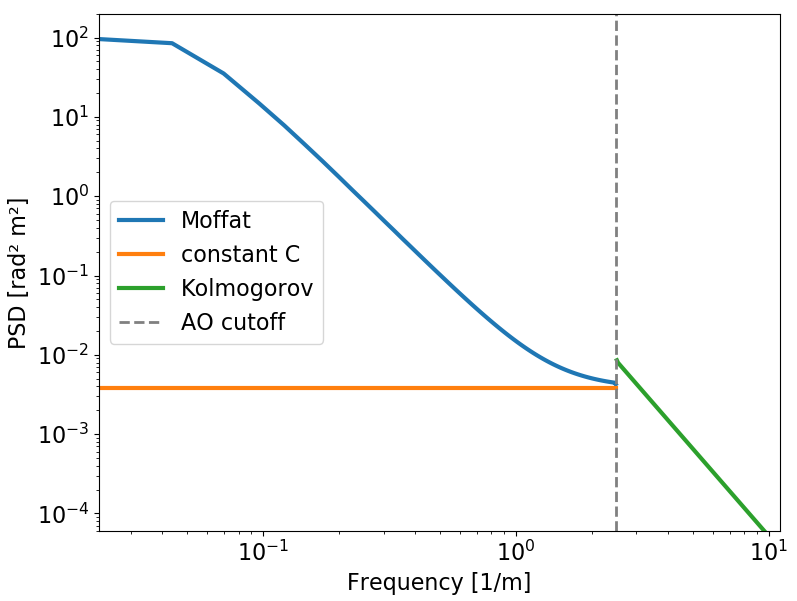}
   \caption{Three components of the PSD model: the Moffat (blue) and the constant contribution (orange) below the AO cutoff frequency, and the Kolmogorov spectrum (green) above the AO cutoff frequency. Discontinuity has been exaggerated by reducing $C$ to show this degree of freedom in our model. Plotting is in logarithmic-logarithmic scale.}
\label{fig:PSD_model}
\end{figure}

Regarding the AO-corrected area, it is difficult to parameterize the phase residual PSD since it depends on the turbulence, the magnitude of the object, the AO loop delay, and the wavefront reconstruction algorithm. Our objective is not to build a full reconstruction of the phase PSD, but to only get a model that can match it. \citet{racine1999speckle} and \citet{jolissaint2002fast} have shown that in extreme AO correction (small residual phase) the shape of the PSF is exactly the shape of the PSD. For partial AO correction, the shapes of the  PSF and PSD are not exactly identical, but are still similar \citep{Fetick2018}. Moreover \citet{Rigaut1998} have shown that the AO residual PSD is the sum of decreasing power laws of the spatial frequency. A Moffat function used in the PSD domain would already describe two regimes due to its shape, one regime for $f\leq\alpha$ and one regime for $\alpha < f < f_\text{AO}$. Adding a constant under the Moffat allows us to describe a third regime near the AO cutoff frequency at $f\lesssim f_\text{AO}$ that is roughly similar to the shape of the aliasing PSD discussed by \citet{Rigaut1998}. All the above pieces of information suggest the possibility of using the Moffat function for a parsimonious parameterization of the AO-corrected PSD, rather than using it to directly parameterize the PSF in the focal plane. The full PSD model is written as
\begin{equation}
\label{eq:psfao}
    W_\phi(\vec{f}) = \left[ \frac{\beta -1}{\pi\alpha_x\alpha_y}\frac{M_A(f_x,f_y)}{1-\left( 1+\frac{f_\text{AO}^2}{\alpha_x\alpha_y} \right)^{1-\beta}} + C \right]_{f\leq f_\text{AO}} + \left[ W_{\phi,\text{Kolmo}}(\vec{f}) \right]_{f>f_\text{AO}}
,\end{equation}
where the Moffat normalization factor ensures a unit integral of the Moffat on the area $f\leq f_\text{AO}$ (see Appendix \ref{sec:appendix_moffat}). Constant $C$ is an AO-corrected phase PSD background. It is useful to model the residual AO PSD near the AO cutoff, where the Moffat function is close to zero. Thus the AO residual phase variance on the circular domain below the AO cutoff frequency is directly
\begin{equation}
\label{eq:sigma2}
    \sigma_\text{AO}^2=A+C\pi f_\text{AO}^2
.\end{equation}
Parameter $A$, added to the $C$ background contribution, has the physical meaning of being the residual variance. An example of our PSD model is given Fig. \ref{fig:PSD_model}. We do not impose continuity at the AO cutoff frequency, so the PSD might be locally discontinuous. Indeed the transition area $f\simeq f_\text{AO}$ between corrected and uncorrected frequencies can lead to strong PSD gradients, which are modelled by an eventual PSD discontinuity.\\

Our PSF model based on the PSD is made of the following set of seven parameters: $\mathcal{S} = \{\alpha_x,\alpha_y,\beta,\theta_R,C,r_0,A\}$ in the asymmetric case.  This reduces to five parameters in the symmetric case (setting $\alpha_y=\alpha_x$ and $\theta_R=0$). Even though symmetric PSFs are sufficient in the majority of cases, asymmetries make it possible to consider PSFs elongated due to strong wind effects or anisoplanetism. Once the parameters $\mathcal{S}$ are set, the PSF is then computed from the AO PSD and static aberrations using Eq. (\ref{eq:PSF}).\\

For the reader interested in deriving the Strehl ratio from our model, we have to compute the integral of the Kolmogorov spectrum above the AO cutoff frequency,
\begin{equation}
\begin{split}
    \sigma^2_\text{halo} & = 0.023r_0^{-5/3} 2\pi\int_{f_\text{AO}}^{+\infty} f^{-11/3}f\text{df}\\
    & = 0.023 \frac{6\pi}{5} (r_0\cdot f_\text{AO})^{-5/3}
\end{split}
.\end{equation}
The Strehl ratio consequently is written as
\begin{equation}
\begin{split}
    \text{S}_\text{R} & = \exp\left[ -(\sigma^2_\text{AO}+\sigma^2_\text{halo})\right] \\
    & = \exp\left[ -A-C\pi f^2_\text{AO}-0.023 \frac{6\pi}{5} (r_0\cdot f_\text{AO})^{-5/3}\right]
\end{split}
.\end{equation}


\section{Validation}
\label{sec:validation}

\subsection{PSF fitting method}
\label{sec:fitting_method}

In this section we deal with images of PSFs (the data) that may come from numerical simulations or observations of stars on VLT instruments. The fitting method consists in finding the PSF parameters so that the model PSF minimizes the square distance to the data PSF:
\begin{equation}
\label{eq:fitting}
    \mathcal{L}(\mathcal{S},\gamma,\zeta,\delta_x,\delta_y) = \sum_{i,j} w_{i,j} \left[ \gamma\cdot h_{i,j}(\mathcal{S},\delta_x,\delta_y)+\zeta - d_{i,j} \right]^2
,\end{equation}
where $h_{i,j}$ is the discretized model of PSF on the pixels $(i,j)$, $\mathcal{S}$ its set of parameters, and $d_{i,j}$ is the data PSF. Since the PSF model (given by Eqs. \ref{eq:PSF} and \ref{eq:psfao}) has a unit flux, it is scaled by $\gamma$ to match the flux of the data PSF, and $\zeta$ accounts for a possible background. The shifts $\delta_x$ and $\delta_y$ centre the PSF with sub-pixel precision on the data (by multiplication of the OTF with the correct phasor). The weighting factor $w_{i,j}$ is the inverse of the noise variance, which takes into account the photon noise and the detector read-out noise. As noted by \citet{mugnier2004}, for high fluxes (typically greater than ten photons per pixel), the Poisson photon noise becomes nearly Gaussian and the weighting factor is written as
\begin{equation}
    w_{i,j} = \frac{1}{\max\{d_{i,j}~,0\}+\sigma_\text{RON}^2}
.\end{equation}
In this case, our approach can be seen from a statistical point of view as maximizing the likelihood of the data $d_{i,j}$ corrupted by photon and read-out noise. We thus minimize $\mathcal{L,}$ which is the neg-logarithm of the likelihood for a Gaussian process.\\

Let us now note that the minimum of $\mathcal{L}$ has an analytic solution for $\gamma$ and $\zeta$ (see Appendix \ref{sec:appendix_flux_bck} for a full demonstration). We actually do not need to numerically minimize over these two parameters, the least-square criterion only relies on the PSF intrinsic parameters $\mathcal{S}$ and the position parameters $(\delta_x,\delta_y)$ as
\begin{equation}
    \mathcal{L}~'(\mathcal{S},\delta_x,\delta_y) = \mathcal{L}(\mathcal{S},\hat{\gamma},\hat{\zeta},\delta_x,\delta_y)
,\end{equation}
where $\hat{\gamma}$ and $\hat{\zeta}$ are the analytic solutions for the flux and the background, respectively. At each iteration of the minimization process, the minimizer evaluates our $\mathcal{L}~'$ criterion with a new set of parameters $(\mathcal{S},\delta_x,\delta_y)$. The current PSF estimate $h(\mathcal{S},\delta_x,\delta_y)$ is computed, then the analytic solutions $\hat{\gamma}$ and $\hat{\zeta}$ are computed. The quantity $\hat{\gamma}\cdot h(\mathcal{S},\delta_x,\delta_y)+\hat{\zeta}$ is used to compute the residuals with the data $d$. Residuals are then provided to the minimizer to estimate a new set of parameters $(\mathcal{S},\delta_x,\delta_y)$.

\subsection{OOMAO end-to-end simulations}
\label{sec:oomao}

The Object-Oriented Matlab Adaptive Optics (OOMAO) toolbox, presented by \citet{OOMAO2014}, provides end-to-end simulations. For each time step, OOMAO generates a turbulent wavefront with a Von-K{\'a}rm{\'a}n spectrum defined as
\begin{equation}
    W_{\phi,\text{VK}}(\vec{f}) = 0.023 r_0^{-5/3} \left[ \left( \frac{1}{L_0} \right) ^2 + f^2 \right] ^{-11/6} 
,\end{equation}
where we have chosen the outer scale $L_0=30$ m. Since $1/L_0\ll f_\text{AO}$, the Von-K{\'a}rm{\'a}n spectrum is consistent with our PSF model using the Kolmogorov spectrum above the AO cutoff frequency. OOMAO then propagates the wavefront through the telescope, simulates the wavefront sensor measurement, performs the wavefront reconstruction, and simulates the mirror wavefront deformation. For each time step, we get a short exposure PSF. Integration over time allows us to retrieve the long-exposure PSF. It is important to notice that the method to compute the PSF is then very different from our model, which directly uses the residual phase PSD in Eq. (\ref{eq:PSF}). We used OOMAO to generate a set of PSFs, corresponding to different wavelengths from $0.5$ um to $2.18$ um, and seven different values of $r_0$ from $7.5$ cm to $25.0$ cm. All $r_0$ are given at $500$ nm. We translate them from the observation wavelength to the reference wavelength of $500$ nm using the theoretical spectral dependency $r_{0,\lambda_1}/r_{0,\lambda_2} = (\lambda_1/\lambda_2)^{6/5}$. For all the PSFs, the telescope parameters are kept unchanged $D=8$ m, $N_\text{act}=32$, sampling at Shannon-Nyquist for all wavelengths. The phase screen consists in one frozen flow (Taylor's hypothesis) turbulent layer translating at $v=10$ m/s. Using these parameters, we generated PSFs corresponding to exposure times of $0.1$, $1$, $10,$ and $100$ seconds. For $0.1$ and $1 $ second, the random OOMAO phase screen did not converge towards a stable state, leading to a strong bias in the $r_0$ estimation. For a $10 $ second exposure PSF, the random fluctuations of the phase are correctly averaged. This is confirmed by the $100$ second exposure PSF, which gives the same $r_0$ estimation as the $10 $ second case. Since a $100$ second exposure is computationally demanding and does not significantly improve the results, we performed all our tests on the $10 $ second exposure time.\\

\begin{table}[!ht]
    \centering
    \begin{tabular}{lcr}
      \hline\hline
      Parameter & Values & Unit\\
      \hline
      Diameter $D$ & 8 & m\\
      $N_\text{act}$ & 32 & --\\
      Wavelength $\lambda$ & 2.18, 1.65, 1.22, 1.0 & \\
                           & 0.85, 0.8, 0.75 ,0.7 &  \\
                           & 0.65, 0.6, 0.55, 0.5 & um \\
      Fried $r_0$ & 7.5, 10.0, 12.5, 15.0 & \\
                  & 17.5, 20.0, 25.0 & cm \\
      Windspeed & 10 & m/s \\
      Outer scale $L_0$ & 30 & m\\
      Exposure & $0.1, 1, 10, 100$ & s\\
      \hline
    \end{tabular}
    \caption{OOMAO parameters summary for our PSF simulations.}
    \label{tab:OOMAOparam}
\end{table}

\begin{table}[!ht]
    \centering
    \begin{tabular}{lcccr}
      \hline\hline
      Param. & Typical range & Lower bound & Guess & Unit\\
      \hline
      $r_0$ & $5-30$ & eps & $18$ & cm\\
      $\alpha_x$ & $10^{-2}-10^{-1}$ & eps & $5\times 10^{-2}$ & m$^{-1}$\\
      $\alpha_y$ & $10^{-2}-10^{-1}$ & eps & $5\times 10^{-2}$ & m$^{-1}$\\
      $\beta$ & $1.1-3$ & 1 + eps & $1.6$ & $-$\\
      $\theta$ & $0-\pi$ & $-$ & $0$ & rad\\
      $C$ & $10^{-3}-10^{-2}$ & $0$ & $10^{-2}$ & rad$^2$m$^2$\\
      $A$ & $10^{-1}-10$ & $0$ & 2 & rad$^2$\\
      \hline
    \end{tabular}
    \caption{Typical range of PSF parameters for OOMAO simulations and SPHERE/ZIMPOL instrument, lower bounds and values used as initial guess for the minimizer. Typical ranges are indicative and may vary according to the considered instrument. The value 'eps' denotes the machine precision. Parameters do not have any upper bound.}
    \label{tab:minimizerparam}
\end{table}

\begin{figure}[!ht]
   \centering
   \includegraphics[width=\columnwidth]{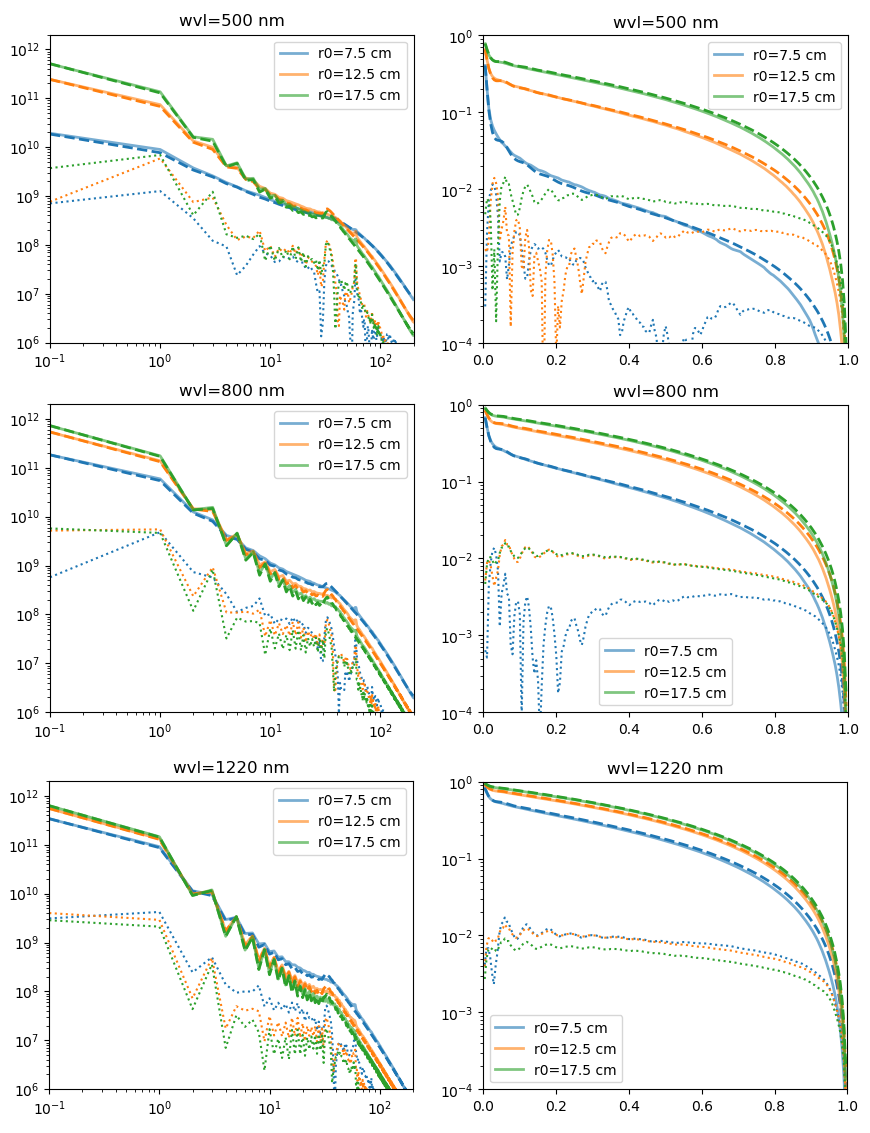}
   \caption{OOMAO PSF fitting with our model. Left: Circular average for PSFs (given in photons). The vertical grey line corresponds to the AO cutoff radius. Right: Corresponding circular average for OTFs (normalized to unity at the null frequency). From top to bottom, three wavelengths are scanned from $500$ nm to $1220$ nm. Colours correspond to three values of the OOMAO required $r_0$. Solid curves: OOMAO. Dashed: fitting. Dotted: residuals. All PSFs, for all wavelengths, are sampled at Shannon-Nyquist.}
\label{fig:OOMAO_PSF}
\end{figure}

Each PSF is then fitted using the $\mathcal{L}~'(\mathcal{S},\delta_x,\delta_y)$ criterion given to an optimizer (e.g. Levenberg-Marquardt, trust region, or Markov chain Monte Carlo). We used the Trust Region Reflective algorithm, called `trf', from the Python/SciPy \citep{scipy} library. This algorithm is gradient based, said to be robust, and allows bounds on the parameters. The robustness of this algorithm was verified for our applications of it to PSF fitting, even though the convexity of the problem is not demonstrated. So far, we have not encountered any local minimum and residuals are always small. For all PSFs, the same initial conditions are provided to the fitting algorithm (see Table \ref{tab:minimizerparam}), in particular we used the same value of $r_0=18$ cm. Using the same initial parameters $\{\mathcal{S},\delta_x,\delta_y\}_\text{init}$ for all fits ensures that our model is suited for minimization procedures and that convergence is ensured even if starting far from the true values. Fitting results are presented on Fig. \ref{fig:OOMAO_PSF}. Our model fits well the OOMAO-generated PSF on both the corrected and the uncorrected area, residuals being on average one to two decades below the PSF. Let us define the relative error between fitted PSF and data PSF
as\begin{equation}
\label{eq:error_rel}
    \epsilon_h = \frac{\sqrt{\sum_{i,j} \left[ \gamma\cdot h_{i,j}(\mathcal{S},\delta_x,\delta_y)+\zeta - d_{i,j} \right]^2}}{\sum_{i,j} d_{i,j}}
.\end{equation}
This error is the L2 norm of the differences between fitting and data, relative to the flux. Considering all the OOMAO fitting, we find an average relative error $\epsilon_h = 6.4\times 10^{-3}$ with a standard deviation of $2.2\times 10^{-3}$. 
For comparison, fitting with a Moffat model gives an average relative error of $\epsilon_h=3.4\times 10^{-2}$ with a standard deviation of $1.6\times 10^{-2}$. Using our model thus increases the fitting accuracy by a factor of approximately $5$ with respect to the former Moffat model of Sect. \ref{sec:moffat}. Regarding the OTFs, the fit is also accurate on the whole frequency range, from low frequencies (mainly the halo) to high frequencies (PSF peak and telescope cutoff). Our model slightly over-estimates frequencies just below the telescope cutoff frequency but this has never been an issue in our applications, such as deconvolution, and is still much better than the Moffat OTF (which has no telescope cutoff frequency and give a poor estimation of the low frequencies).\\

Regarding the flux, we consider the relative error between the flux $\hat{\gamma}$ analytically estimated by our model fitting method, and the OOMAO flux that is directly the sum of the data on all the pixels:
\begin{equation}
    \epsilon_\gamma = \frac{\hat{\gamma}-\sum_{i,j}d_{i,j}}{\sum_{i,j}d_{i,j}}
.\end{equation}
On all our OOMAO simulations, we find an average relative error of $-1.96\%$, indicating a small underestimation of the flux with our fitting method. The standard deviation of this relative error is $1.11\%$, and the range of variation is $[-3.43\%,2.77\%]$.

\subsubsection{Fried parameter $r_0$ estimation}

\begin{figure}[!ht]
   \centering
   \includegraphics[width=\columnwidth]{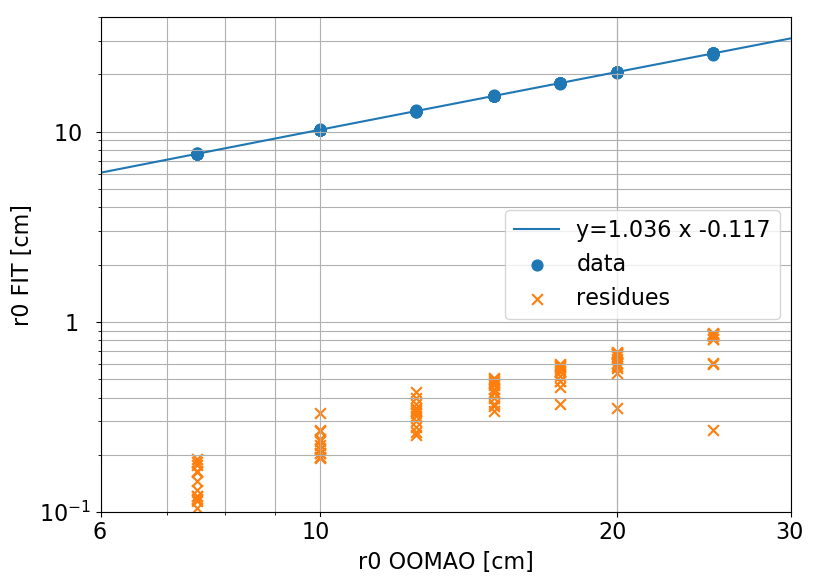}
   \caption{Fried parameter $r_0$ estimated by fitting versus the $r_0$ used in OOMAO to generate the PSF. All $r_0$ are given at $500$ nm. Here are shown results on 84 different PSFs, corresponding to seven values of $r_0$ and 12 different wavelengths. The line is the linear fit between our $r_0$ estimation and OOMAO $r_0$. Crosses show residuals $|r_{0,\text{FIT}}-r_{0,\text{OOMAO}}|$. A log-log scale is used to show on the same graph both data and residuals.}
\label{fig:OOMAO_r0}
\end{figure}

As shown in Fig. \ref{fig:OOMAO_r0}, our $r_0$ estimation is consistent with the OOMAO value of $r_0$. We find the best linear fit
to be\begin{equation}
\label{eq:OOMAOr0r0fit}
    r_{0,\text{FIT}} = 1.038~r_{0,\text{OOMAO}} - 0.132
,\end{equation}
where values are given in centimetres. The Pearson correlation coefficient is $\mathcal{C}_\text{Pearson} = \text{Cov}(r_{0,\text{FIT}},r_{0,\text{OOMAO}})/\sqrt{\text{Var}(r_{0,\text{FIT}})\cdot \text{Var}(r_{0,\text{OOMAO}})} = 0.99992$. This result fully confirms our $r_0$ estimation with respect to OOMAO simulations with sub-centimetre precision.

\subsubsection{AO residual variance $\sigma^2_\text{AO}$ estimation}

\begin{figure}[!ht]
   \centering
   \includegraphics[width=.9\columnwidth]{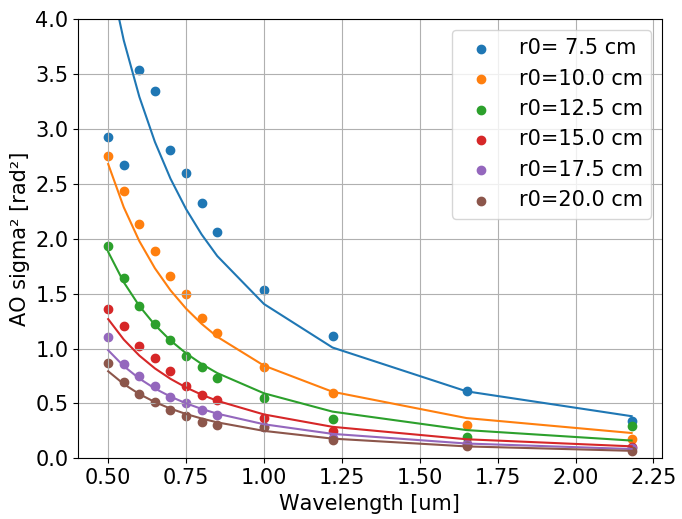}
   \caption{Estimation of the $\sigma^2_{AO}$ from PSF fitting versus the wavelength (dots). Colours correspond to the seven different values of OOMAO $r_0$. Curves of parametric equations $\sigma^2=a\lambda^{-2}$ are fitted on the data.}
\label{fig:OOMAO_sigma2}
\end{figure}

Theoretically our model should also be able to retrieve the residual variance $\sigma^2_{AO}$ on the corrected area and follow a $\lambda^{-2}$ power law. Figure \ref{fig:OOMAO_sigma2} shows the fitting estimation of this variance versus the wavelength. This data is then fitted with curves of equation $a\lambda^{-2}$. Except the two outliers for minimal $r_0=7.5$cm at low wavelength ($\lambda\simeq 500$nm), the $\lambda^{-2}$ power law is a good estimation of the $\sigma^2_\text{AO}$ evolution. This result gives confidence in the estimated parameter. Data from the real time computer (RTC) could be used in the future to provide the $\sigma^2_\text{AO}$ parameter for PSF estimation. The $\lambda^{-2}$ spectral dependence is also an asset to shift the PSF from one wavelength to another.

\subsubsection{Constant $C$ estimation}

\begin{figure}[!ht]
   \centering
   \includegraphics[width=.9\columnwidth]{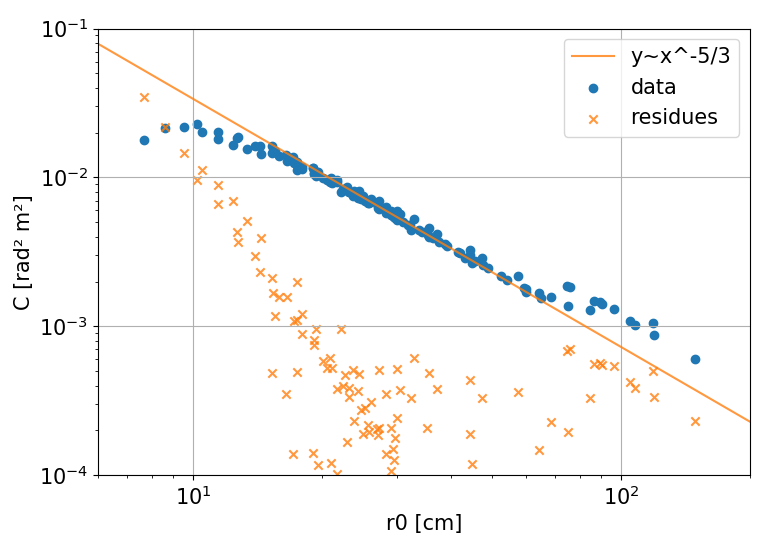}
   \caption{Estimation of the PSF constant $C$ versus the $r_0$ given at the observed wavelength (dots). A $r_0^{-5/3}$ fitting equation (solid line) is applied on the data. Residuals between each data point and the $r_0^{-5/3}$ power law are also shown (crosses).}
\label{fig:OOMAO_aliasing}
\end{figure}

The $C$ term in Eq. (\ref{eq:psfao}) accounts for multiple sources of residual PSD, including wavefront aliasing and other AO residual errors. Since this constant is dominated by the Moffat PSD in the core, it becomes more important near the AO cutoff, where the aliasing dominates. Since the aliasing scales in $r_0^{-5/3}$ \citep{Rigaut1998}, we look for similar $r_0$ dependencies for the PSF constant $C$. Figure \ref{fig:OOMAO_aliasing} shows a clear decrease of $C$ with $r_0$. Fitting the estimated $C$ with a $r_0^{-5/3}$ power law shows a good match, with small residuals for nearly all $r_0$ values. The power law is not exactly $-5/3\simeq -1.67$ but is closer to $-1.46$ for this OOMAO case. However, one can still think about normalizing the constant in Eq. (\ref{eq:psfao}) by
\begin{equation}
    C = C' r_0^{-5/3}
\end{equation}
and perform fitting over $C'$ instead of $C$. This would reduce the variation range of this parameter. Reducing bounds or standard deviation of a parameter is an asset for constraining the model and improving minimization processes.

\subsection{High performance imager ZIMPOL}
\label{sec:ZIMPOL}

The Spectro Polarimetric High-contrast Exoplanet REsearch (SPHERE) instrument \citep{beuzit2008sphere,beuzit2019sphere} of the VLT includes the powerful SPHERE Adaptive optics for eXoplanets Observation (SAXO) system described in \citet{fusco2014final} and \citet{sauvage2010saxo}. The AO real time computer is built on the ESO system called SPARTA \citep{Fedrigo2006sparta}, which stands for Standard Platform for Adaptive optics Real Time Applications. In particular, for each observation, SPARTA is able to give an estimate of $r_0$ from the mirror voltages and wavefront sensor slopes. \\

\begin{figure}[!ht]
   \centering
   \includegraphics[width=\columnwidth]{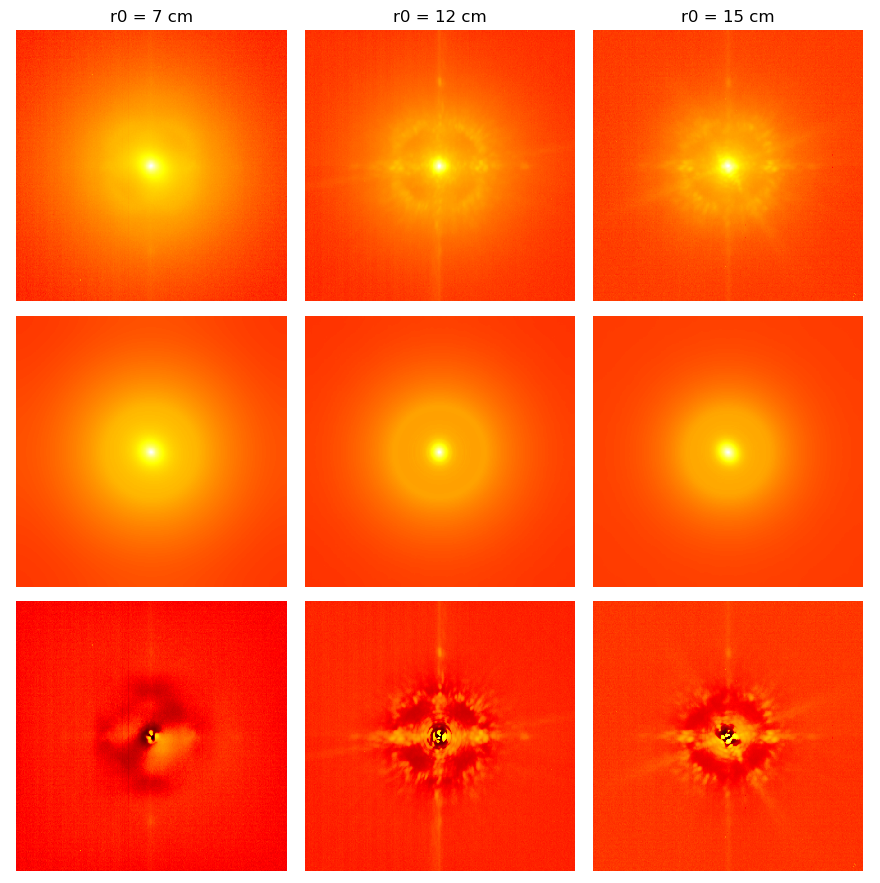}
   \caption{Three ZIMPOL PSFs (top), model fittings (middle), residuals (bottom). Left: Minimal $r_0$ of the sample. Middle: Median $r_0$. Right: Maximal $r_0$. The main differences are due to some static aberrations not taken into account in our model (only the pupil and its central obstruction are taken into account). The hyperbolic arcsine of the intensity is shown to enhance details. The same intensity scale is used per column (data, model, residuals), but differs between columns.}
\label{fig:ZIMPOL_3_psf}
\end{figure}

The Zurich IMaging POLarimeter (ZIMPOL) instrument \citep{Schmid2018} is mounted at the focal plane of SPHERE. ZIMPOL is also used as a very efficient imager at visible wavelengths. One of its applications in non-coronagraphic mode is the observation of asteroids \citep{LP2018Vernazza,LP2018Matti,Fetick2019} as part of an ESO Large Program (ID 199.C-0074, PI P. Vernazza). PSFs from stars were observed with the ZIMPOL N\_R filter (central wavelength 645.9 nm, width 56.7 nm) during the Large Program. When PSFs are saved together with the SPARTA telemetry, we are able to correlate $r_0$ given by our fitting and $r_0$ given by SPARTA. In our sample, 28 PSFs were saved along with the SPARTA telemetry. These PSFs were obtained during different nights, on stars of different magnitudes, with various seeing conditions. We fitted these 28 PSFs with our PSF model. Figure \ref{fig:ZIMPOL_3_psf} shows three of the 28 fittings, for the smallest $r_0$ of the sample, the median $r_0$, and the largest $r_0,$ respectively. Our fitted PSFs match the shape of the core and halo. The average of relative error defined in Eq. (\ref{eq:error_rel}) is $\epsilon_h= 2.5 \times 10^{-3}$, with a standard deviation of $1.2\times 10^{-3}$. We only consider diffraction due to the telescope 8 m aperture and its central obstruction in the static OTF $\tilde{\mathrm{h}}_T$ (see Eq. (\ref{eq:PSF})). Consequently non-circularly-symmetric effects, such as the spiders or static aberrations visible on Fig. \ref{fig:ZIMPOL_3_psf}, are not modelled. Even if the spiders could have been included in our model, we have deliberately chosen to ignore them in the $\tilde{h}_T$ term since they are negligible in comparison to the other dominant effects (AO residual core, turbulent halo, 8 m aperture diffraction). When performing PSF fitting, the contribution of these effects not taken into account in $\tilde{h}_T$ might bias the atmospheric term $\tilde{h}_A$ during fitting procedure and slightly offset the estimation of the $\mathcal{S}$ parameters. Moreover these ZIMPOL images are field stabilized, meaning rotating spiders, which are harder to model. Pupil stabilized images would make the description of the spider diffraction effect easier.\\

\begin{figure}[!ht]
   \centering
   \includegraphics[width=.95\columnwidth]{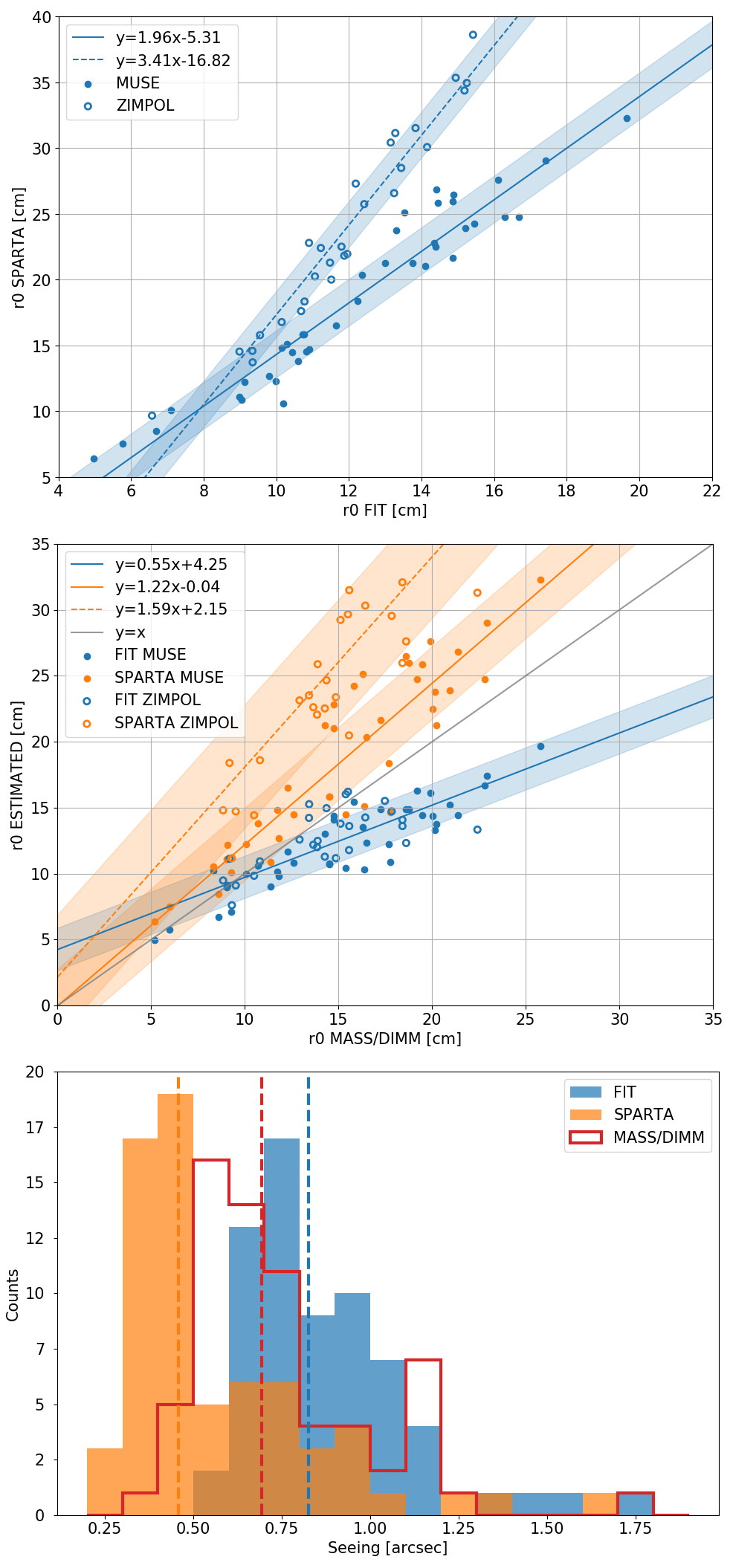}
   \caption{Zenital $r_0$ at 500 nm estimated on MUSE (filled circles, 40 data points) and SPHERE/ZIMPOL (empty circles, 27 data points) using three methods: PSF fitting, SPARTA, and MASS/DIMM. Top: $r_{0,\text{SPARTA}}$ versus $r_{0,\text{FIT}}$. A different linear tendency is found for MUSE (plain line) and SPHERE/ZIMPOL (dashed line). Shaded areas show the standard deviation between data points and the best linear fit. Middle: $r_{0,\text{SPARTA}}$ and $r_{0,\text{FIT}}$ versus $r_{0,\text{MASS/DIMM}}$. Two linear tendencies are identified for SPARTA, and only one for PSF fitting. Bottom: Histograms of seeing estimated with the three methods on both instruments. Dashed vertical lines show median values of $0.46"$ (SPARTA), $0.69"$ (MASS/DIMM), and $0.83"$ (PSF fitting).}
\label{fig:ZIMPOL_r0}
\end{figure}

The top graph on Fig. \ref{fig:ZIMPOL_r0} shows that the values estimated by SPARTA (median = 22 cm) are greater than the values given by fitting (median = 13 cm). The best linear fit between $r_0$ estimated  by fitting and SPARTA is
\begin{equation}
\label{eq:r0r0fit}
    r_{0,\text{SPARTA}} = 3.41~r_{0,\text{FIT}} - 16.82
.\end{equation}
The Pearson correlation coefficient between the two series $r_{0,\text{FIT}}$ and $r_{0,\text{SPARTA}}$ is $\mathcal{C}_\text{Pearson}= \text{Cov}(r_{0,\text{FIT}},r_{0,\text{SPARTA}})/\sqrt{\text{Var}(r_{0,\text{FIT}})\cdot \text{Var}(r_{0,\text{SPARTA}})} =0.97$. From this data it appears that the estimates of $r_{0,\text{FIT}}$ and $r_{0,\text{SPARTA}}$ are not identical, however they show a strong correlation. We further investigated the difference between the SPARTA and the fitting estimates thanks to the ESO atmospherical monitoring using the Multi-Aperture Scintillation Sensor (MASS) combined with the Differential Image Motion Monitor (DIMM). Since the MASS/DIMM instrument is located apart from the telescopes, it does not see exactly the same turbulent volume as the telescopes and does not suffer the same dome effect. There might be some uncertainties between MASS/DIMM $r_0$ estimations and telescope $r_0$ estimations (PSF fitting or RTC) due to the spatial evolution of the turbulence. Nevertheless this instrument is a valuable indicator of the Paranal atmospheric statistics. For each PSF observation we retrieved the associated MASS/DIMM seeing estimation within a delay of $\pm 3$ minutes (see Fig. \ref{fig:ZIMPOL_r0}, middle and bottom graphics). The median seeing estimated by the MASS/DIMM is $0.69"$, to be compared with a median seeing of $0.46"$ for SPARTA and $0.83"$ for PSF fitting. The over-estimation of the SPARTA $r_0$ with respect to the MASS/DIMM $r_0$ has been already discussed by \citet{Milli2017}. The exact origin of the difference between these three estimations has not be found. Nevertheless we note that estimations with SPARTA are based on RTC measurements of the low spatial frequencies of the phase (sensitive to the Von-K{\'a}rm{\'a}n outer scale $L_0$), whereas our fitting method is based on the PSF halo corresponding to the high spatial frequencies. Our PSF fitting method might be sensitive to telescope internal wavefront errors if they have not been previously calibrated and taken into account in the PSF model.\\ 

Even if there is still an uncertainty on the true value of $r_0$, the strong correlation between SPARTA and fitting estimations is sufficient for many applications. Indeed it is still possible to get $r_{0,\text{SPARTA}}$ from telemetry, use Eq. (\ref{eq:r0r0fit}) to translate it into $r_{0,\text{FIT}}$ , and get an estimate of the PSF halo. This method constrains the model for future PSF estimations without having access to the actual image of the PSF.\\

\subsection{MUSE integral field spectrograph}
\label{sec:muse}

The Multi-Unit Spectroscopic Explorer MUSE \citep{bacon2006probing,bacon2010muse} is an integral field spectrograph (IFS) working mainly in the visible, from $\sim 465$ nm to $\sim 930$ nm. MUSE is equipped with the Ground Atmospheric Layer Adaptive Optics for Spectroscopic Imaging (GALACSI) adaptive optics system \citep{Stroebele2012} to improve its spatial resolution in two different modes, the so-called narrow-field mode (NFM) and wide-field mode (WFM), to correct different sizes of field of view. The AO facility uses four laser guide stars (LGS) \citep{calia2014lgs} to perform a tomographic reconstruction of the turbulent phase. A $589$ nm dichroic is present in the optical path to avoid light contamination from the sodium AO lasers, so no scientific information is available around this wavelength. Let us also note that MUSE is undersampled (sampling is 25 mas in NFM) on the whole available spectrum. Our PSF model manages the undersampling issue by oversampling the PSD and the OTF to safely perform numerical computations. The given PSF is then spatially binned to retrieve the correct sampling. The shape of the PSF can be retrieved, but this method has lower precision on the parameters' estimation inherent to undersampled data. These differences between the MUSE instrument and SPHERE/ZIMPOL allow us to test the versatility of our PSF model. Additionally the spectral resolution is an asset to validate our model at different wavelengths and to study the spectral evolution of our PSF parameters.\\

\begin{figure}[!ht]
   \centering
   \includegraphics[width=\columnwidth]{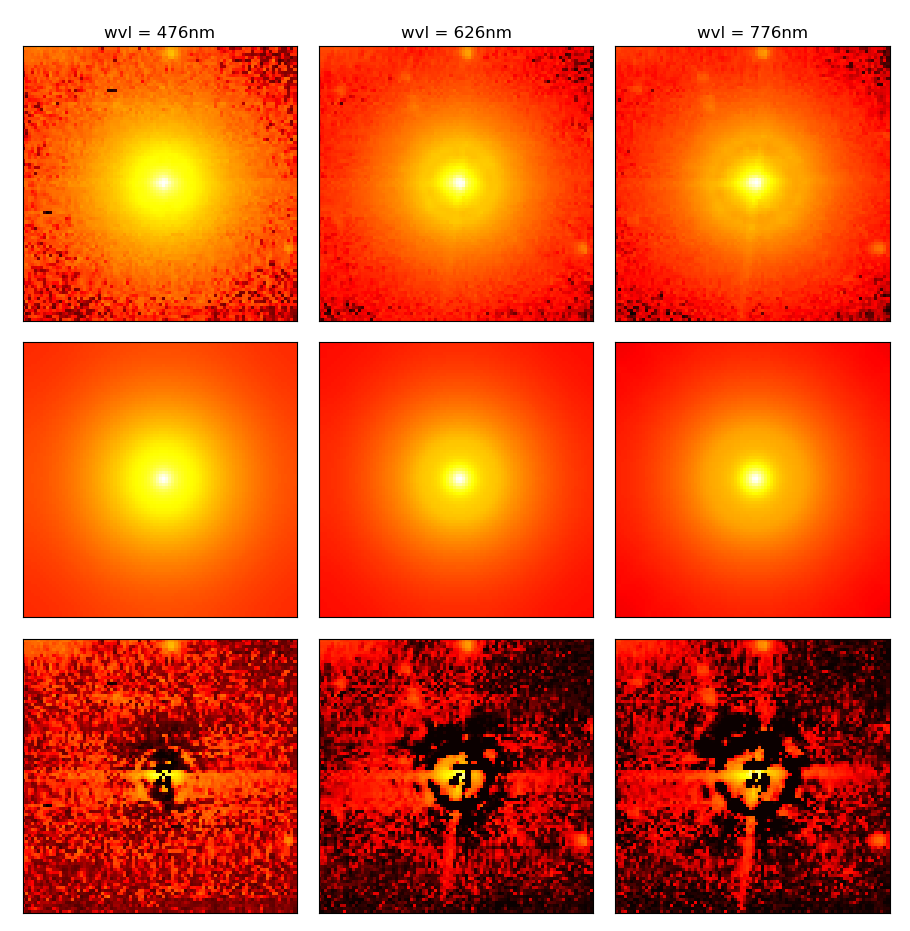}
   \caption{MUSE PSFs (top), fittings (middle), and residuals (bottom) of the same star at three different wavelengths over the 92 spectral bins actually fitted. The hyperbolic arcsine of the intensity is shown to enhance details.}
\label{fig:muse_psf}
\end{figure}

During the May-June 2018 commissioning phase, MUSE observed multiple targets in narrow-field mode. Among these targets, we have access to 40 PSFs observed on different stars, during different nights and at different seeing conditions. These selected PSFs have been spectrally binned into 92 bins of 5nm each to increase the signal to noise ratio and reduce the number of fittings. Then fitting is performed independently, spectral bin by spectral bin, without any spectral information on the targets or the atmosphere. Figure \ref{fig:muse_psf} shows one MUSE datacube PSF fitting at three different wavelengths. The evolution of the AO correction radius is clearly visible in both the data and the model. As for ZIMPOL, we did not take into account static PSF (except the occulted pupil diffraction), which is the main visible difference between data and model. Fainter stars visible in the field did not affect the fitting and appear clearly in the residuals. For the 40 datacubes PSF, the relative error is $\epsilon_h=3.3\times 10^{-3}$. This result is similar to the previous ones on OOMAO and SPHERE/ZIMPOL. Secondary stars in the field also count in the residual error computation.\\

\begin{figure}[!ht]
   \centering
   \includegraphics[width=.9\columnwidth]{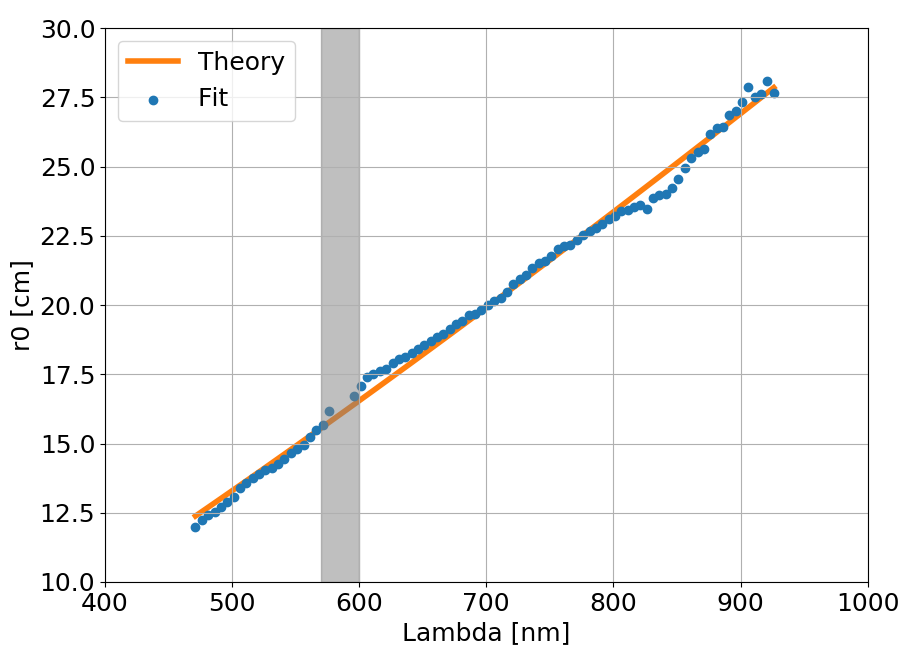}
   \caption{Estimation of the $r0$ by fitting of the PSF MUSE at 92 different wavelengths (blue dots), and comparison with a theoretical law in $\lambda^{6/5}$ (orange line). The best match between data and theory is achieved for $r_0=13.3$ cm. The grey area corresponds to missing wavelengths due to the sodium notch filter.}
\label{fig:muse_r0}
\end{figure}

The evolution of $r_0$ with the wavelength is shown on Fig. \ref{fig:muse_r0} for one datacube PSF. A least square fitting between our data points and the theoretical $\lambda^{6/5}$ evolution of the $r_0$ gives a spectral averaged estimation of $\overline{r_0}=13.3$ cm at $500$ nm. Our fitted $r_0$ matches well the theory, with a standard deviation of $0.3$ cm.\\

So far each spectral bin is fitted independently, however the spectral deterministic trend we recover is an asset for PSF determination. It makes possible the fitting of the whole datacube with only one $r_0$ parameter. The statistical contrast -- ratio of the number of measurements over the number of unknowns -- would be increased. It would improve fitting robustness, especially for faint stars where the halo is strongly affected by noise.\\

The results of fitting on the 40 datacubes give statistical information on the $r_0$ estimation. As for the SPHERE/ZIMPOL case, we have access to SPARTA and MASS/DIMM data to correlate with our fitting estimations (see Fig. \ref{fig:ZIMPOL_r0}). The Pearson correlation coefficient between PSF fitting and SPARTA is $\mathcal{C}_\text{Pearson}=0.96$, which is similar to the SPHERE/ZIMPOL case. However, we get the linear relationship
\begin{equation}
\label{eq:r0r0fit_muse}
    r_{0,\text{SPARTA}} = 1.96~r_{0,\text{FIT}} - 5.31
,\end{equation}
which is different from the SPHERE/ZIMPOL (Eq. (\ref{eq:r0r0fit})). The exact origin of this different trend is unknown, nevertheless let us note that the actual implementation of the phase PSD estimation is slightly different for the SPHERE and for the MUSE instruments. For SPHERE the Von-K{\'a}rm{\'a}n outer scale $L_0$ is set to $25$ m, whereas for MUSE the $L_0$ is estimated jointly with $r_0$. This SPARTA double trend is corroborated by MASS/DIMM information (Fig. \ref{fig:ZIMPOL_r0}, middle graph). On the other hand, $r_0$ estimation using our PSF model gives similar results on both SPHERE/ZIMPOL and MUSE instruments. This confirms the robustness of our PSF fitting method.


\section{Conclusions}
\label{sec:ccl}

In this article we developed a parametric model of long-exposure AO-corrected PSF. The particularity of this model is to parameterize the phase PSD using a Moffat core and a turbulent Kolmogorov halo. This model also incorporates prior knowledge of the telescope, such as the optical cutoff frequency, the obstruction and spider shapes, and even the static aberrations if they are calibrated, for example by phase diversity \citep{Mugnier2008}. This model only requires five parameters for circularly symmetrical PSFs, and seven for asymmetrical ones. The sparsity of this PSF model makes it suitable for numerical computation, such as minimization algorithms or least-square fits. Tests on both simulated and real data validated the appropriateness of our model.\\

One substantial advantage of our model over focal plane models is to use physical parameters such as the Fried parameter $r_0$ and the residual AO variance $\sigma^2_\text{AO}$. Since these parameters are physical, their values in our PSF model can be correlated to external measurements. Tests on both OOMAO simulations and on-sky data (from the SPHERE/ZIMPOL and MUSE instruments) confirmed the physical meaning of the $r_0$ parameter used in our PSF. The ultimate goal would be to only use physical parameters in the PSF description.\\

Our model has already shown usability for different seeings and different instruments, with different AO-correction quality. This shows the robustness and versatility of the model. We also plan to use it to parameterize the PSF for the future instruments on bigger telescopes such as the Extremely Large Telescope (ELT).\\

Finally, the small number of parameters makes this model suited for image post-processing techniques such as deconvolution  of long-exposure images. Deconvolution using parametric PSFs has already been demonstrated by \citet{Drummond1998} and \citet{Fetick2019}. We plan to develop a myopic deconvolution algorithm estimating both the observed object and the PSF parameters in a marginal approach similar to \citet{Blanco2011}.


\begin{acknowledgements}
    This work was supported by the French Direction G{\'e}n{\'e}rale de l'Armement (DGA) and Aix-Marseille Universit{\'e} (AMU). This work was supported by the Action Sp{\'e}cifique Haute R{\'e}solution Angulaire (ASHRA) of CNRS/INSU co-funded by CNES. This project has received funding from the European Union's Horizon 2020 research and innovation program under grant agreement No 730890. This material reflects only the authors views and the Commission is not liable for any use that may be made of the information contained therein. This study has been partly funded by the French Aerospace Lab (ONERA) in the frame of the VASCO Research Project.\\
    
    The authors are thankful to Pierre Vernazza for providing data related to his ESO Large Program ID 199.C-0074.
\end{acknowledgements}

\bibliographystyle{aa.bst} 
\bibliography{references.bib}


\begin{appendix}
\section{Integral of a truncated Moffat}
\label{sec:appendix_moffat}

\begin{figure}[!ht]
   \centering
   \includegraphics[width=.6\columnwidth]{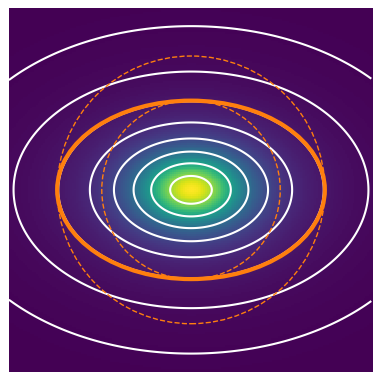}
   \caption{Visualization of an elongated Moffat (colour map), its elliptical level curves (white), and the ellipse $\mathbb{E}(R_x,R_y)$ inside which the integral is computed. Dashed circles of radius $R_x$ and $R_y$ help to visualize the error done when computing the integral over the ellipse instead of a circle. A very elongated Moffat is shown here ($\alpha_x=1.5\alpha_y$).}
\label{fig:moffat_ellipse}
\end{figure}

A Moffat function as given in Eq. (\ref{eq:moffat}) shows elliptical contours (see Fig. \ref{fig:moffat_ellipse}). In this appendix we compute the integral of the Moffat function inside one of these elliptical contours, called $\mathbb{E}(R_x,R_y)$, of semi-major axis $R_x$ and semi-minor axis $R_y= (\alpha_y/\alpha_x) R_x$. The integral to calculate is
\begin{equation}
    I(R_x,R_y) = \iint_{\mathbb{E}(R_x,R_y)} M_A(x,y) \mathrm{dx}\mathrm{dy}
.\end{equation}
Let us perform the change of variables
\begin{equation}
\phi : 
\left\lbrace
\begin{array}{cl}
\mathbb{R}_+^* \times [-\pi,\pi[ & \longrightarrow \mathbb{R}\times\mathbb{R} \backslash (0,0) \\
(r,\theta) & \longrightarrow (\alpha_x r \cos\theta,\alpha_y r\sin\theta)\\
\end{array}
\right.
.\end{equation}
The determinant of the Jacobian is
\begin{equation}
\det J_\phi = \begin{vmatrix}
\alpha_x\cos\theta & -r\alpha_x\sin\theta\\
\alpha_y\sin\theta & r\alpha_y\cos\theta
\end{vmatrix}
=\alpha_x\alpha_y r
.\end{equation}
The integral of the Moffat in the ellipse is rewritten as
\begin{equation}
\begin{split}
I(R_x,R_y) & = \int_{-\pi}^{\pi}\int_0^{R_x/\alpha_x} |\det J_\phi|~  M_A(\phi(r,\theta)) \mathrm{dr}\mathrm{d}\theta\\
 & = 2\pi A\alpha_x\alpha_y\int_0^{R_x/\alpha_x}  [1+r^2]^{-\beta}\\
 & = A\frac{\pi\alpha_x\alpha_y}{\beta-1}\left\{ 1-\left[ 1+(R_x/\alpha_x)^2 \right]^{1-\beta}\right\}
\end{split}
,\end{equation}
where we assumed $\beta\neq 1$.\\

In the nearly circular regime $\alpha_x\simeq\alpha_y$ and $R_x\simeq R_y$ so the integral over the ellipse $\mathbb{E}(R_x,R_y)$ is nearly equal to the energy in a disk of radius $R_x$. This assumption is made for our model of PSD to compute the residual variance below the AO cutoff frequency in Eq. (\ref{eq:psfao}).\\

Moreover it is possible to calculate the integral of the Moffat function over the whole plane. The assumption $\beta>1$ is mandatory to get a bounded integral (finite energy). Under this assumption, letting $R_x\to +\infty$ and $R_y\to +\infty,$ one finds
\begin{equation}
    I(+\infty,+\infty) = A\frac{\pi\alpha_x\alpha_y}{\beta-1}
.\end{equation}
Consequently, in order to get a Moffat of unit integral over the whole plane, one must choose the Moffat amplitude factor as
\begin{equation}
    A = \frac{\beta-1}{\pi\alpha_x\alpha_y}
.\end{equation}

\section{Analytic solution for the flux and background}
\label{sec:appendix_flux_bck}

The minimum of $\mathcal{L}$ (given in Eq. \ref{eq:fitting}) has an analytic solution for $\gamma$ and $\zeta$ since nulling the partial derivative of $\mathcal{L}$ towards these parameters gives
\begin{equation}
\begin{split}
\frac{\partial\mathcal{L}}{\partial\gamma}=0 &  \Longleftrightarrow \sum_{i,j} w_{i,j} h_{i,j} \left[ \gamma\cdot h_{i,j}+\zeta - d_{i,j} \right]=0\\
&  \Longleftrightarrow \gamma \sum_{i,j} w_{i,j} h_{i,j}^2 + \zeta  \sum_{i,j} w_{i,j} h_{i,j} = \sum_{i,j} w_{i,j} h_{i,j}d_{i,j}
\end{split}
\end{equation}
and
\begin{equation}
\begin{split}
\frac{\partial\mathcal{L}}{\partial\zeta}=0 &  \Longleftrightarrow \sum_{i,j} w_{i,j} \left[ \gamma\cdot h_{i,j}+\zeta - d_{i,j} \right]=0\\
&  \Longleftrightarrow \gamma \sum_{i,j} w_{i,j} h_{i,j} + \zeta  \sum_{i,j} w_{i,j} = \sum_{i,j} w_{i,j} d_{i,j}
\end{split}
.\end{equation}
These two equations are linear in $\gamma$ and $\zeta$. They can be written within the matrix formalism
\begin{equation}
    \mathcal{A}\cdot
    \begin{pmatrix}
    \zeta \\ \gamma\\
    \end{pmatrix}
    =
    \sum_{i,j} w_{i,j} d_{i,j}
    \begin{pmatrix}
    1 \\ h_{i,j}\\
    \end{pmatrix}
,\end{equation}
where $\mathcal{A}$ is the $2\times 2$ matrix of the system defined as
\begin{equation}
\mathcal{A} = \sum_{i,j} w_{i,j}
\begin{pmatrix}
1 & h_{i,j}\\
h_{i,j} & h_{i,j}^2\\
\end{pmatrix}
.\end{equation}
In order to invert $\mathcal{A}$, we need to make sure that $\text{det}(\mathcal{A})\neq 0$. The determinant of $\mathcal{A}$ is
\begin{equation}
\begin{split}
\text{det}(\mathcal{A}) & = \left( \sum_{i,j} w_{i,j}\right) \left( \sum_{i,j} w_{i,j}h_{i,j}^2\right) - \left( \sum_{i,j} w_{i,j}h_{i,j}\right)^2\\
 & = \left( \sum_{i,j} \sqrt{w_{i,j}}^2\right) \left( \sum_{i,j} (\sqrt{w_{i,j}}h_{i,j})^2\right) - \left( \sum_{i,j} (\sqrt{w_{i,j}})(\sqrt{w_{i,j}}h_{i,j})\right)^2
\end{split}
.\end{equation}
We can now define the two vectors $\sqrt{W}=\{\sqrt{w_{i,j}}\}$ and $\sqrt{W}H=\{\sqrt{w_{i,j}}h_{i,j}\}$. Using these notations the determinant is rewritten as
\begin{equation}
     \text{det}(\mathcal{A}) = \Vert \sqrt{W} \Vert^2 \Vert\sqrt{W}H\Vert^2-\vert\langle \sqrt{W},\sqrt{W}H \rangle\vert^2
.\end{equation}
We recognize the Cauchy-Schwarz inequality. It follows that $\text{det}(\mathcal{A})\geq 0$, with equality only if $\sqrt{W}$ and $\sqrt{W}H$ are colinear vectors. The colinearity is written as
\begin{equation}
    \begin{split}
        \sqrt{W}\parallelsum\sqrt{W}H & \Longleftrightarrow \exists k\in\mathrm{R} \mid \forall (i,j), k\sqrt{w_{i,j}} = \sqrt{w_{i,j}} h_{i,j}\\
         & \Longleftrightarrow \exists k\in\mathrm{R} \mid \forall (i,j)\text{ where }w_{i,j}\neq 0,~ k =  h_{i,j}
    \end{split}
.\end{equation}
This states that the determinant of $\mathcal{A}$ is null only if the PSF model $h$ is constant on each point where $w_{i,j}\neq 0$. Since our PSF is not constant on the domain $w_{i,j}\neq 0$, we ensure that $\text{det}(\mathcal{A})\neq 0$ and the analytic solution for $\gamma$ and $\zeta$ is written as
\begin{equation}
    \begin{pmatrix}
    \zeta \\ \gamma\\
    \end{pmatrix}
    =
    \mathcal{A}^{-1}
    \sum_{i,j} w_{i,j} d_{i,j}
    \begin{pmatrix}
    1 \\ h_{i,j}\\
    \end{pmatrix}
.\end{equation}

\end{appendix}

\end{document}